\documentclass[conference]{IEEEtran}
\IEEEoverridecommandlockouts
\usepackage{cite}
\usepackage{amsmath,amssymb,amsfonts}
\usepackage{algorithmic}
\usepackage{graphicx}
\usepackage{subcaption}
\usepackage{multirow}
\usepackage{multicol}
\usepackage{float}
\usepackage{textcomp}
\usepackage{xcolor}
\usepackage{color}
\usepackage{booktabs} 
\usepackage{makecell}
\usepackage{enumitem}
\usepackage{etoolbox}
\usepackage{titlecaps}
\usepackage{hyperref}
\hypersetup{
    colorlinks=true,      
    linkcolor=blue,
    filecolor=blue,      
    urlcolor=blue,
    citecolor=cyan,     
}

\def\BibTeX{{\rm B\kern-.05em{\sc i\kern-.025em b}\kern-.08em
    T\kern-.1667em\lower.7ex\hbox{E}\kern-.125emX}}

\begin{document}
\title{TEAdapter: Supply Vivid Guidance for \\ Controllable Text-to-Music Generation\\
{\footnotesize \textsuperscript{}}
\thanks{$^{\dag}$ Equal contribution.}\\
\thanks{$^{\ddag}$ Corresponding author.}
}


\makeatletter
\newcommand{\linebreakand}{%
  \end{@IEEEauthorhalign}
  \hfill\mbox{}\par
  \mbox{}\hfill\begin{@IEEEauthorhalign}
}
\makeatother

\author{\IEEEauthorblockN{Jialing Zou$^{\dag}$}
\IEEEauthorblockA{\textit{East China Normal University}\\
Shanghai, China\\
51255901103@stu.ecnu.edu.cn}
\and
\IEEEauthorblockN{Jiahao Mei$^{\dag}$}
\IEEEauthorblockA{\textit{East China Normal University} \\
Shanghai, China\\
10215102440@stu.ecnu.edu.cn}
\and
\IEEEauthorblockN{XuDong Nan}
\IEEEauthorblockA{\textit{East China Normal University} \\
Shanghai, China\\
71255901084@stu.ecnu.edu.cn}
\linebreakand
\IEEEauthorblockN{Jinghua Li}
\IEEEauthorblockA{\textit{East China Normal University}\\
Shanghai, China\\
10215102415@stu.ecnu.edu.cn}
\and
\IEEEauthorblockN{Daoguo Dong$^{\ddag}$}
\IEEEauthorblockA{\textit{East China Normal University}\\
Shanghai, China\\
dgdong@cs.ecnu.edu.cn}
\and
\IEEEauthorblockN{Liang He}
\IEEEauthorblockA{\textit{East China Normal University}\\
Shanghai, China\\
lhe@cs.ecnu.edu.cn}
}

\maketitle
\begin{abstract}
Although current text-guided music generation technology can cope with simple creative scenarios, achieving fine-grained control over individual text-modality conditions remains challenging as user demands become more intricate. Accordingly, we introduce the \textbf{TEA}cher \textbf{Adapter} (\textbf{TEAdapter}), a compact plugin designed to guide the generation process with diverse control information provided by users. In addition, we explore the controllable generation of extended music by leveraging TEAdapter control groups trained on data of distinct structural functionalities. In general, we consider controls over global, elemental, and structural levels. Experimental results demonstrate that the proposed TEAdapter enables multiple precise controls and ensures high-quality music generation. Our module is also lightweight and transferable to any diffusion model architecture. Available code and demos will be found soon at \href{https://github.com/Ashley1101/TEAdapter}{https://github.com/Ashley1101/TEAdapter}.
\end{abstract}


\vspace{-5pt}
\section{Introduction}
\label{sec:intro}

Music has the ability to comfort the soul. When a user hears a deeply moving piece of music, they often become immersed in it, 
hoping create similar tunes by themselves.  Cross-modal music generation technology has brought all brand new hope to turn this expectation into reality. Existing text-to-music technologies attempt to generate music based on textual labels, phrases, or detailed descriptions, but current solutions still encounter several issues as follows.

Non-professionals often lack Specialized knowledge of musical theory, thus tend to use conceptual vocabulary to express their interest, which may result in unsatisfactory generated samples due to insufficient details. Moreover, there is a modal gap between text and music, so inevitable loss of information will appear during the process of modal conversion. Last but not least, the present generation methods still exhibit numerous deficiencies in control aspects, making it challenging to support more refined and personalized creative practice. 

In fact, it is natural to consider utilizing existing music as a carrier of users' demands, as an individual's perception of desired music often originates from their past listening experiences. These mature compositions can serve as references, offering insights into rhythm signatures, melodic contours, chord structures and instrument timbres. In this case, the three types of cruxes aforementioned can be simultaneously solved or at least alleviated. To achieve enhanced precision in control, we consider providing targeted designs from three different levels: global, element, and structure.

Inspired by the field of image generation, extra control can be delivered through additional modules attaching to existing generative models, such as ControlNet\cite{zhang2023adding} and T2I-Adapter\cite{mou2023t2i}. Therefore, we intend to extract specific information based on the user-provided ``teacher music'', feeding it to corresponding additional module, in order to promote the performance of generated music on a certain aspect. Besides, we will also incorporate some global tags into previous text conditions as new constraints for generation.

Furthermore, generating extended music composition is another noteworthy task, as the musical expressiveness and structure transformation within a short music segment are limited. Although existing text-to-music models are capable of generating authentic short music clips, they often fail in generating high-quality long music. For example, applying MusicGen\cite{copet2023musicgen} or AudioLDM\cite{liu2023audioldm} to generate long music will result in highly repetitive looped music in terms of auditory perception.  Therefore, we propose a novel framework based on \textbf{TEAdapter} to generate structurally complete long music. The core idea of this approach is to utilize multiple \textbf{TEAdapter sets} to generate different music structural parts and
 seamlessly merge them into a complete composition using inpainting techniques.

In summary, the principle contributions of this article are listed as follows:

\begin{itemize}
\item We propose a novel control enhancement plugin, \textbf{TEA}cher \textbf{Adapter} (\textbf{TEAdapter}), which facilitates controls spanning multiple categories, scales, and input forms, and is easily adaptable to any other diffusion-based music generation models.
\vspace{-1px}
\item Our model enables the creation of extended music with comprehensive musical narrative structure, allowing control over each different section. Controllability is able to cover the entire composition by delivering music inpainting schemes.
\vspace{-1px}
\item Our training approach is resource-efficient, enhancing performance while minimizing training costs. Objective evaluations affirm that our method achieves competitive performance across diverse metrics.
\end{itemize}

\section{Related Work}
\vspace{0.06in}
\subsection{Automated Music Generation}
When the generation target is symbolic music (in the form of piano-roll, MIDI, etc.), the original task can be converted into a sequence prediction problem. \cite{liu2022symphony} designs a novel Multi-track Multi-instrument Repeatable (MMR) representation for symphonic music modeling, applying a Transformer-based auto-regressive language model with specific 3D positional embeddings. \cite{liang2020pirhdy} suggest a framework named PiRhDy to represent pitch and rhythm in musical note events as tokens, ensuring a balanced generation of both melody and accompaniment. Nevertheless, the compositions created in this way rely on digital libraries later, consequently sounding mechanical and monotonic. 

When the generation target is audio signals (in the form of waveform, spectrogram, etc.), it is capable to inject more dynamic performance effects into musical works. \cite{huang2023noise2music} introduces Noise2Music, which utilizes a cascaded diffusion model to produce 24kHz music segments lasting 30 seconds. However, the lack of constraints on musical theory rules may lead to inferior outcomes.

\vspace{-5px}
\subsection{Fine-grained Controllable Generation}
Nowadays, the primary methods of music generation relies on textual conditions, but real-world applications demand more complex and precise control. In the field of image generation, ControlNet\cite{zhang2023adding} duplicates the encoder of UNet during the denoising process of Stable Diffusion\cite{rombach2022high}, and influences the original generation by fine-tuning the parameters of the "trainable copy" part. Similarly, T2I-adapter\cite{mou2023t2i} freezes the weights of the Stable Diffusion\cite{rombach2022high} also, then builds a more lightweight adapter module comprising of several residual and down-sampling blocks instead. Uni-ControlNet\cite{zhao2023uni} further tackles the issue of excessive network size by classifying conditions into local and global categories, adopting different addition and training methods.

Meanwhile, there is limited exploration in the same direction of music generation. \cite{wang2020learning} raises a decoupling scheme for polyphonic chords and textures. It indicates that these two interpretable elements can serve as universal features for controllable music generation. Music ControlNet\cite{wu2023music}, inspired by ControlNet\cite{zhang2023adding}, provides guidance in melody, dynamics, and rhythm. Nevertheless, it involves relatively high training costs.
\section{Methodology}
\label{sec:method}
\vspace{0.06in}
Our TEAdapter is capable to work in any diffusion-based architecture to enhance controllability in music generation. In this paper, we conduct experiments on one of the existing state-of-the-art unified audio generation framework, AudioLDM 2. Furthermore, we construct several TEAdapter-groups to establish distinct musical narrative structures. The subsequent sections offer an in-depth overview of how TEAdapter employs various control types at different levels. Additionally, we elaborate on adapter composition, control information extraction and guidance integration process. The complete pipeline of our work is displayed in Figure~\ref{fig:pipeline}.
\subsection{Preliminary: AudioLDM 2}
AudioLDM 2\cite{liu2023audioldm2}, a versatile audio generation model, facilitates cross-modal audio generation encompassing sound effect, music, and speech. Leveraging a special \textit{Language of Audio (LOA)} as an intermediary bridge between target and conditional modalities, AudioLDM 2 enables self-supervised pretraining without text annotations given in advance. Multiple types of conditions can be encoded by pretrained Large Language Models (LLMs) to acquire LOA, and ultimately applied in the denoising process of diffusion models.
Despite AudioLDM 2 holds huge generation potential, it encounters challenges like poor structure in generating music, excessive focus on specific vocabulary that leads to neglect of other words, and insufficient diversity in the results. Consequently, we decide to evaluate the effectiveness of TEAdapters according to the improvement in controllability of AudioLDM 2.
\subsection{Method: TEAdapter}
\subsubsection{Problem Formulation}
Taking AudioLDM 2 as an illustration, considering the input raw audio as $x$. In the self-supervised training process of AudioLDM 2, the initial step involves constructing LOA: $Y=A (x)$, where $A$ denotes the AudioMAE encoder\cite{huang2022masked}. Meanwhile, $x$ will be compressed into a hidden vector as $z_0$ through VAE encoder. The denoising process of T-UNet predicts the added noise $\epsilon$ through $\hat{\epsilon_t}=U (z_t, t, Y)$ for each step, where $z_t$ means the latent representation of step $t$. During inference, the large language model $M$ should be fine-tuned first to obtain $\hat{Y}=M (C)$ that approximates the ground truth $Y$. After that, LOA features can be simulated based on condition $C$ in the subsequent sampling process.

\begin{figure*}[t]
\vspace{0.06in}
    \centering
    \includegraphics[width=0.90\linewidth]{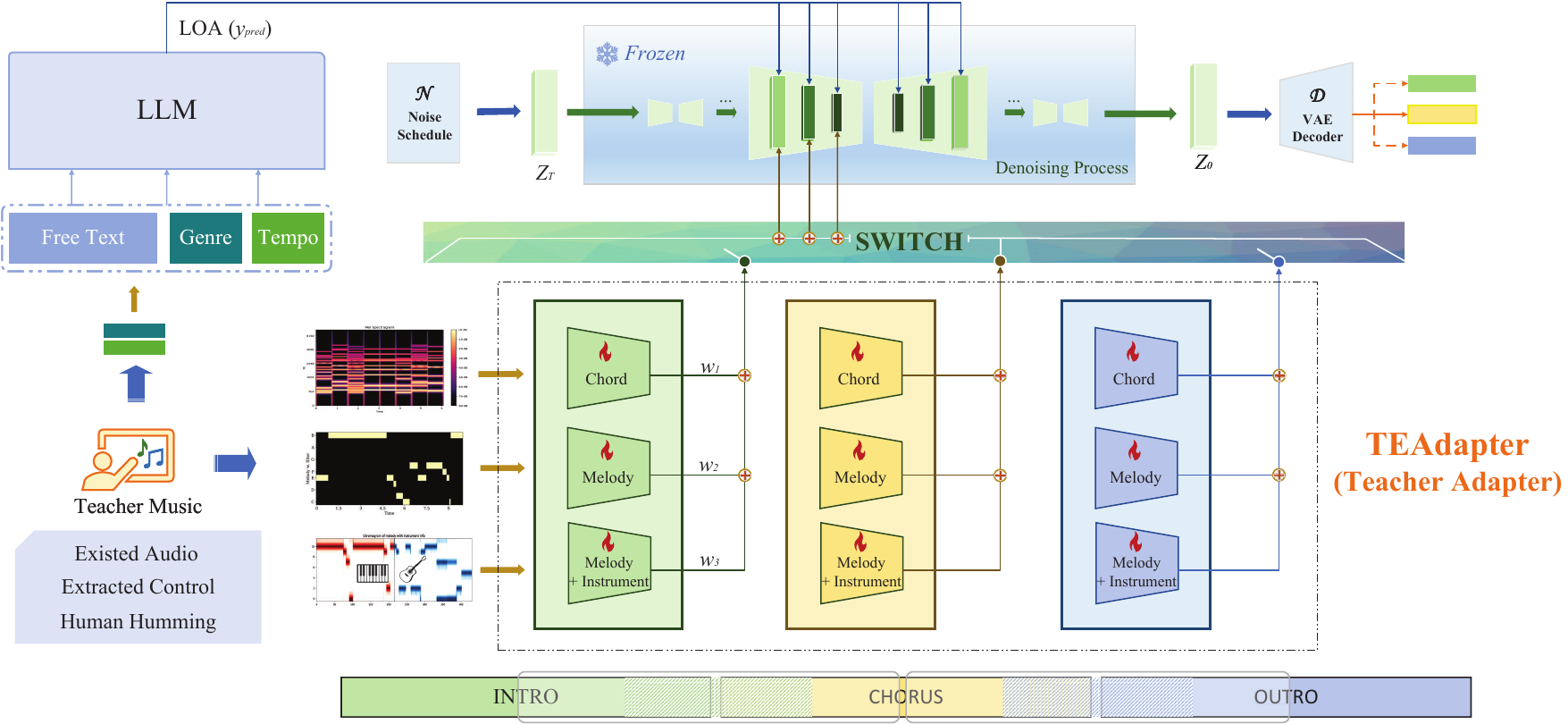}%
    \caption{The overall architecture comprises a frozen music diffusion model and multiple TEAdapter groups. The music diffusion model accepts free text with additional global labels as inputs. Each TEAdapter receives different features extracted from ``teacher music'', and can be combined into a group with corresponding weight parameter ($\omega$). The switch section determines which TEAdapter-group is used. The schematic at the bottom illustrates inpainting process. The white sliding window represents the input for each inpainting iteration, and the patches indicate the re-generated portions.}
    \label{fig:pipeline}
    \vspace{-10pt}
\end{figure*}

Keeping the parameters of pretrained AudioLDM 2 model frozen, TEAdapter is designed to augment control by fusing additional guidance into the denoising process. Following the configuration of the T2I-adapter\cite{mou2023t2i}, each TEAdapter is fabricated by alternating residual convolutional blocks and down-sample layers. In the training period, chord progression and primary melody are separated from initial music and transformed into the same audio format. Then take properties of musical instrument with melody into account, we will derive three types of control in total, each will be utilized for training one corresponding TEAdapter. The central role of our TEAdapter is to extract features $Y_{ec}$ of distinct scales from the current additional control $C_{ec}$, and sum up with the intermediate output $U_t=\{{U_t}^1, {U_t}^2,..., {U_t}^n\}$ of UNet encoder to impact the generation process, where $n$ is the number of feature maps generated at the current time step $t$. 

At the stage of inference, users are informed to upload one or more ``teacher music'' recorded as $G=\{g_1, g_2,...\}$ and select desired guidance categories for each clip, determining a set of extra control conditions $C_{ec}=\{{C_{ec}}^1, {C_{ec}}^2,... {C_{ec}}^l\}$, where $l$ refers to the number of conditions. At this point, the updated denoising process will be:
\begin{equation}
z_{t-1}=z_t- \hat{\epsilon_t},~\hat{{\epsilon_t}}=U(z_t, t, Y, Y_{ec};\theta)
\end{equation}

In the formula, ${Y_{{ec}^i}}=\{y^1_{{ec}^i}, y^2_{{ec}^i},...,y^n_{{ec}^i}\}$ signifies the TEAdapter features extracted from ${C^i_{ec}}$, and $y^j_{{ec}^i}$ is the feature extracted by the i-th condition after passing through the j-th ResNet block of the TEAdapter. Users can set the corresponding weight $w_i$ to any $C^i_{ec}$, and the result $Y$ of the weighted sum of all control condition codes will be added to each UNet feature map:
\begin{equation}
Y_{ec}=\sum \limits_{i}^l(w_i * Y^i_{ec})
\end{equation}

In a summary, we can optimize the training goal on the basis of loss function below:
\begin{equation}
L_{AD}=\mathbb{E}_{z_0,t,Y,Y_{ec},\epsilon \sim N(0,1)}\Big[\Arrowvert \epsilon-U(z_t, t, Y, Y_e;\theta) \Arrowvert\Big]
\end{equation}

\subsubsection{Elemental Control}
Elemental controls consist of features changing over time, including chord progression,primary melody, and instrument timbre. Chords determine the overall auditory tone, serving as accompaniment that complements the main melody but remains relatively independent. Melody, the part most likely to be perceived and remembered by users, reflects the most notable discrepancy across various music types. Thus either changes of melodic or arrangement can lead to distinct auditory sensations.

\begin{figure}[t]
    \centering
\includegraphics[width=0.92\linewidth]{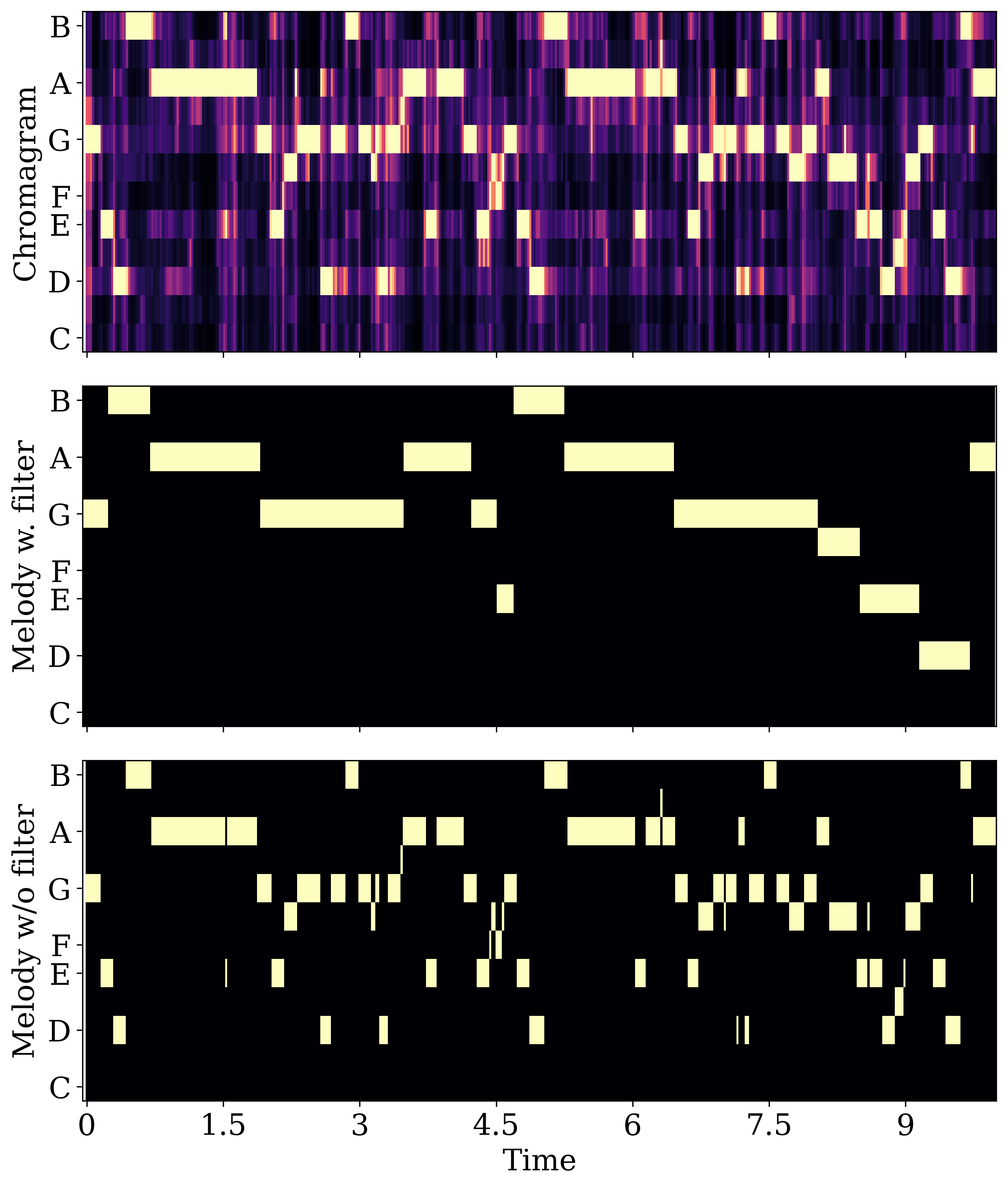}%
    \caption{Visualization of (1) the original Chromagram on a 10-second pop music clip, (2) corresponding melody extraction result with filter operation and (3) without filter operation.}
  \label{fig:melody_extraction}
  \vspace{-10pt}
\end{figure}

Furthermore, switching instruments can evoke diverse emotional responses from the audiences. For instance, the violin's tone evokes melancholy and elegance, while the ukulele imparts a light and bright sensation. Instead of relying on sole instrument category label, our TEAdapter further considers the situation that different instruments play in the same generation process.

\begin{figure*}[t]
    \centering
    \includegraphics[width=0.95\linewidth]{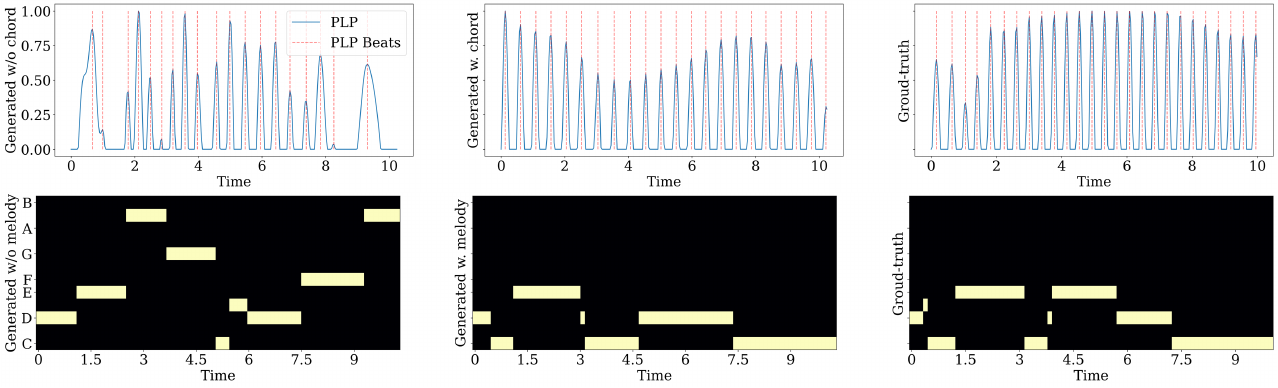}%
    \caption{ (1) The red dashed line represents the positions of beats estimated by PLP. It can be observed that employing chord guidance yields better rhythm stability compared to not using chord guidance.
(2) Compared to not using melody guidance, the melody of the results generated with melody guidance is closer to the reference music. }
    \label{fig:tempo_melody}
\end{figure*}

\textbf{Chord Progression}. To extract chord information from audio or MIDI files, chord analysis tools\footnote{\url{https://github.com/p3zo/chord-progressions/}} available on GitHub has been selected. When dealing with audio files, the first step is to obtain the corresponding spectral data based on the original audio, involving performing operations such as peak detection and HPCP feature extraction. The next step is to identify rhythm of the audio files through the RhythmExtractor2013 tool in the Essentia library\footnote{\url{https://essentia.upf.edu/}}. Based on the position and loudness of beats, the music can be divided into neat sections. Finally, the most prominent pitches within each bar will be mapped to musical scales of the greatest probability, identifying the overall chord progression of the music.

  
    
  

\textbf{Primary Melody}. We employ the librosa\cite{librosa2023} library to extract primary melody. In each chromagram, the horizontal axis represents time, the vertical axis reflects pitches of 12 semitones(see subplot 1 of Figure~\ref{fig:melody_extraction}). Applying smooth filtering and argmax operations on each time frame, we will gain one-hot encoding of primary melody from source music. To prevent model overfitting, we set the hop size to a large value of 2048 to drop redundant melody details. The main steps of smooth filtering involve: apply harmonic-percussive source separation to eliminate transients (like drums), Nearest-neighbor smoothing to eliminate passing tones and sparse noise \cite{cho2011feature} and local median filtering to suppress remaining discontinuities.(extraction results are shown in Figure~\ref{fig:melody_extraction} and generation results are shown in Figure~\ref{fig:tempo_melody}).

\textbf{Musical Instruments}. We utilize EfficientAT\cite{schmid2023efficient} model fine-tuned on the OpenMic\cite{humphrey2018openmic} instrument dataset to predict multiple non-overlapping instrument labels for every music. Subsequently , an sound font file that is most similar to the instrument's timbre is selected to play the aforementioned chord or  melody.

\subsubsection{Structural Control}
Music possesses temporal structural elements, encompassing segmental similarities and distinct functional components. Structural integrity serves as a crucial distinction between complete compositions and solely segments.

To compose long music with structural characteristics, we employ several TEAdapter-groups to generate different music components, simplifying music structure into three categories: intro, chorus and outro based on the annotation method proposed by the Salami\cite{smith2011design} dataset. Subsequently, we utilize the music structure analysis tool MSAF\cite{nieto2016systematic} to categorize the music dataset FMA\cite{fma_dataset} into aforementioned three subsets employed to train separate TEAdapter-groups to generate different structural parts of music. However, directly concatenating these music segments cannot preserve perceptual consistency at the junction parts. Therefore, we implement music inpainting technology discussed in AudioLDM\cite{liu2023audioldm}. The whole process(shown at the bottom of Figure~\ref{fig:pipeline}) involves adding a mask at the junction parts of two music segments, re-generating the masked portion of the music, and finally obtaining the complete music with improved perceptual consistency. An example can be found in Figure~\ref{fig:inpt}.
\subsubsection{Global Control}
Moreover, to implement global music control through mode and tempo restrictions, we opt to incorporate them directly as suffixes to free text. This approach aims to further standardize the original text conditions.
\section{EXPERIMENT}

\begin{figure*}[t]
\vspace{-5pt}
    \centering
    \includegraphics[width=0.95\linewidth]{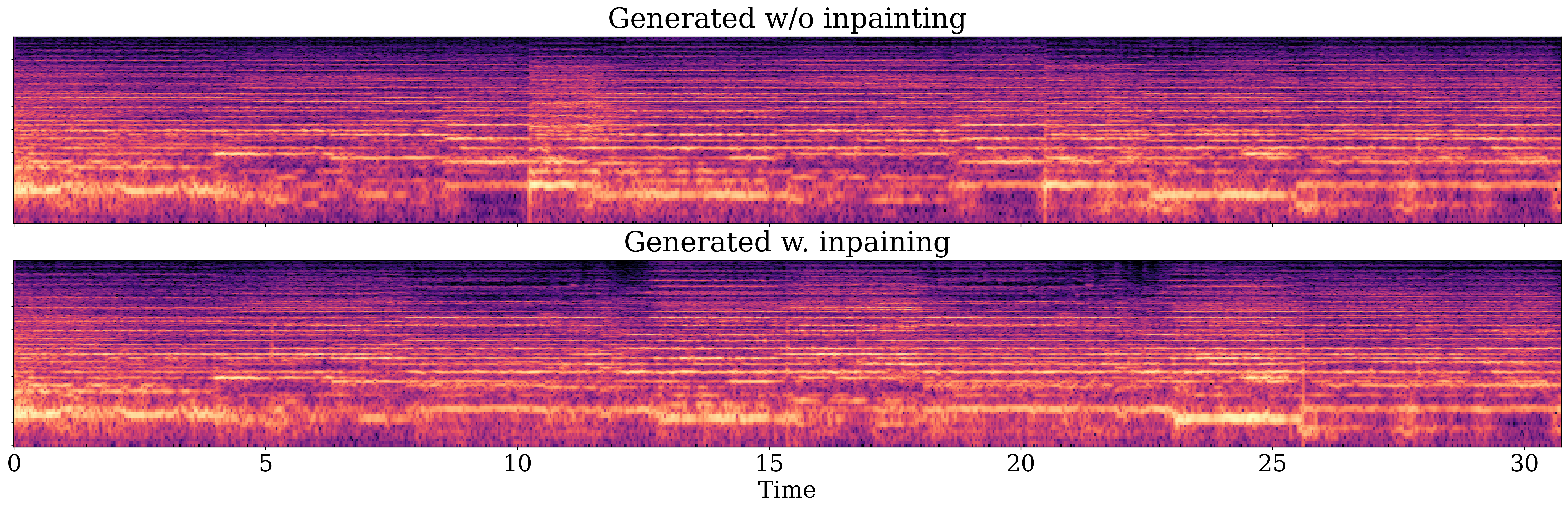}%
    \vspace{-10pt}
    \caption{ Spectrograms visualization of the structural control generated music. Noticeable splicing artifacts can be observed at the 10- and 20-second positions in the first subplot. In contrast, the inpainting result exhibits improved continuity and consistency.}
    \label{fig:inpt}
    \vspace{-10pt}
\end{figure*}

\vspace{-5px}
\label{sec:experiment}
\vspace{0.06in}
\subsection{Experimental Setup}
\textbf{Dataset}. We collect data from MusicCaps\cite{agostinelli2023musiclm} and FMA\cite{fma_dataset} datasets for training. MusicCaps comprises 5.5K 10-second clips annotated by professional musicians, including a 1K subset that is well-proportioned across genres. We employ the unbalanced set of MusicCaps for training and validation. FMA contains 106K 30-second musical segments without captions, so prompts like ``generate a \{genre\} music" are used as alternatives. Adopting the music structure analysis tool MSAF\cite{nieto2016systematic}, we cut each sample into 10-second piece of data referring to three categories: intro, chorus, and outro. Then we randomly select 20K pieces from each subclass for joint training with the MusicCaps training set. To assess the bonus brought by our modules on the general performance of original model and the guiding power of individual TEAdapters, we report uniform metrics of evaluation on the genre-balanced subset of MusicCaps. Moreover, we conduct separate training on the three subsets of FMA to further guarantee structural control.

\textbf{Model Specifics}. Basic architecture of our TEAdapter follows the fundamental design of T2I-adapter\cite{mou2023t2i}, comprising four feature extraction blocks and three downsampling blocks. The initial input, representing music with dimensions $\Large[1024,64,1\Large]$, undergoes a pixel unshuffle operation to be scaled into $256*16$. Next, feature maps with scales of $\Large\{32*2, 64*4, 128*8, 256*16 \Large\}$ will be yielded to align with the corresponding output of the UNet encoder in AudioLDM 2\cite{liu2023audioldm2}. We choose the pretrained AudioLDM 2 as our baseline model. Additionally, both the ``teacher music'' and samples from dataset are uniformly resampled to 16kHz. The training utilizes an Adam optimizer with the learning rate set to $1e-5$. Both at training and inference stages, the diffusion model sampling involves 200 steps with a batch size of 16.

\textbf{Evaluation Metrics}. To determine the degree of control enhancement, we devise two metrics: melody accuracy and beat stability. Melody accuracy assesses whether the pitch classes extracted from generated music match with the reference music, and beat stability measures the variance of intervals between adjacent downbeats obtained by the Predominant Local Pulse (PLP) \cite{grosche2010extracting}estimation. For evaluating overall text-to-music performance, we leverage Fréchet Audio Distance (FAD)\cite{kilgour2018fr}, Kullback-Leibler Divergence (KL)\cite{koutini2021efficient}, and CLAP score (CLAP)\cite{wu2023large} on MusicCaps testset. FAD gauges the audibility of music, KL divergence quantifies the distribution discrepancy between generated music and source data, and the CLAP score reflects the alignment between produced results and text condition.

\subsection{Experimental Results}
\textbf{Global generation capability}. Table~\ref{tab:baselines} shows the comparison of our method with AudioLDM\cite{liu2023audioldm}, MusicGEN\cite{copet2023musicgen} and AudioLDM 2\cite{liu2023audioldm2}. We reproduce AudioLDM by method published on its official GitHub website\footnote{\url{https://github.com/haoheliu/AudioLDM-training-finetuning/}}, download the official no-melody-guided edition($1.5B$) of MusicGEN and \textit{audioldm2-music-665k} checkpoint of AudioLDM 2. For our approaches, we test both on single TEAdapters and TEAdapter-groups. The former includes three types of TEAdapters with chord, pure melody and melody with instrument information guidance (corresponding to \textbf{Chord-Adapter}, \textbf{Melody-Adapter} and \textbf{Melody+Instrument-Adapter} in turn), while the latter integrates two groups of modules \{Chord-Adapter, Melody-Adapter\} and \{Chord Adapter, Melody+Instrument-Adapter\} respectively, so that acquires two versions of TEAdapter-groups, i.e. \textbf{TEAdapter-CM} and \textbf{TEAdapter-CMI}. According to our validation, we set the control weights of TEAdapters in each group to 0.4 and 0.6 respectively.
The final results reveal our method's superior performance in FAD and CLAP scores, while demonstrating comparable performance in the KL-divergence. Specifically, the Chord-Adapter excels in FAD, underscoring the significant role of chords in comprehensive auditory experience. The TEAdapter-CM outperforms other models in CLAP score, highlighting the influence of both melody and chord in constituting core musical semantics. The injection of instrument information results in a bit decrease in CLAP, probably because that our instrument labels can not overlap in time intervals, which will be optimized in our future work. Additionally, our method slightly trails behind AudioLDM 2 in KL divergence, a trend also observed in MusicGEN. We guess that incorporating extra guidance may contribute to enhancing diversity within the original distribution. Furthermore, the integration of different TEAdapters seems to be less than fully effective, and we will investigate potential solutions in future research.

\textbf{Performance of Controllability}. As is depicted in Table~\ref{tab:commands} and Figure~\ref{fig:tempo_melody}, the results illustrates that our model beat the original model in melody and instrument accuracy. It is indicated that the corresponding guidance enhances elemental control in generated music, and opens avenues for music rearrangement and style transfer based on ``teacher music''. Despite AudioLDM 2 excels in average beat stability, it exhibits higher variance than our model. Figure~\ref{fig:tempo_melody}(1) illustrates a representative case where guidance from ``teacher music'' significantly contributes to reinforcing generation stability.

\textbf{Long audio generation}. We train TEAdapter on FMA-intro, FMA-chorus, and FMA-outro subsets which represent different musical functions. As the example shown in Figure~\ref{fig:inpt}, we dig out melody and instrument information from three given ``teacher music'', samples generated by distinct TEAdapters but based on the same text description are similar as a whole but have certain differences. We can splice it together and apply the inpainting technology to smooth the connection, for the sake of producing a complete 30-second long music.

\begin{table}[t]
\vspace{-15pt}
  \centering
  \caption{Comparison of Generation Quality on the Balanced Subset of MusicCaps}
  \label{tab:baselines}
  \begin{tabular}{c|ccc}
    \toprule
    \hline
    & \textbf{FAD~$\downarrow$} & \textbf{KL~$\downarrow$} & \textbf{CLAP(\%)} \\
    \midrule
    AudioLDM\cite{liu2023audioldm}&13.89 &3.72 &21.2\\
    MusicGEN\cite{copet2023musicgen} &4.80 &1.54 & -\\
    AudioLDM 2\cite{liu2023audioldm2}&5.24 &\textbf{1.49} &25.9\\
    \midrule
    Chord-Adapter &\textbf{3.87} &1.53 &25.9\\
    Melody-Adapter &4.17 &1.51 &26.0\\
    Melody+Instrument-Adapter &5.74 &1.55&25.9\\
    \midrule
    TEAdapter-CM &4.28 &1.52 &\textbf{26.7}\\
    TEAdapter-CMI &4.55 &1.52 &26.5\\
    \bottomrule
  \end{tabular}
  \vspace{-6px}
\end{table}

\begin{table}[t]
  \centering
  \caption{Comparison of Controllability on the Balanced Subset of MusicCaps}
  \label{tab:commands}
  \begin{tabular}{c|cccc}
    \toprule
    \hline
     & \textbf{~~Melody~~} & \multicolumn{2}{c}{\textbf{Beat}} & {\textbf{Inst.}}\\
      & \multirow{2}*{Acc.(\%)} & \multicolumn{2}{c}{Stab.} & \multirow{2}*{Acc.(\%)}\\
    & & Mean~$\downarrow$ & Std~$\downarrow$ & \\
    \midrule
    AudioLDM 2\cite{liu2023audioldm2}&24.1 &\textbf{.060} &.073 &29.8 \\
    \midrule
    Chord-Adapter &24.0 &.066 &\textbf{.068} & 31.0 \\
    Melody-Adapter &\textbf{32.9} &.065 &.069 & 31.7\\
    \midrule
    TEAdapter-CM &27.1 &.064 &.072 & 29.7\\
    TEAdapter-CMI &27.8 &.065 &.073 &\textbf{31.8}\\
    \bottomrule
  \end{tabular}
  \vspace{-6px}
\end{table}

  
    


\section{Conclusion}
\label{conclusion}
\vspace{0.04in}
In this paper, we proposed a new scheme of controllable music generation, supplying fine-tuned control of the generation through a small component called TEAdapter. Our method achieved a balance between stability and diversity of various aspects including melody, instrument, rhythm and so on. We further expanded on our work through complementary annotations, music inpainting and other strategies. Experimental results and comparisons with the latest state-of-the-art models in music generation manifested that our approach was versatile and cost-effective, and capable of reaching excellent performance. We commit to actively addressing the current deficiencies of our approach in the follow-up work.
\section{Acknowledgement}
\label{sec:acknowledgement}
This work was supported by the National Key R\&D Program of China (2022ZD0161800), and the computation is performed in ECNU Multifunctional Platform for Innovation(001).

%

\small{
\bibliographystyle{IEEEbib}
\bibliography{total}
}

\end{document}


\sloppy

\section*{Supplementary material}

\setcounter{figure}{0}
\renewcommand*{\thefigure}{A\arabic{figure}}
\setcounter{table}{0}
\renewcommand*{\thetable}{A\arabic{table}}

This supplementary material will provide more detail of our work,
which are organized as outlined below:

\begin{itemize}

\item Section \ref{A}: Further details regarding the model employed in the FVS.

\item Section \ref{B}: An augmented description of the details involved in preparing training data during the adaptation of the model. This includes a detailed account of the iterative process for filtering training data, the detailed content of the prepared prompts, and examples of high-quality images selected using the FVS. Additionally, it include the implementation of the adapting process.

\item Section \ref{C}: Further analysis of the correlation of FVS and CLIP Score and the adapted model, encompassing supplemental analysis of examples presented in the article as well as additional analytical results.

\end{itemize}


\section{More details of the model in FVS}
\label{A}
In the Aesthetic Assessment section of the FVS, after the extraction of various image attributes, we apply a feature selection approach to integrate these attribute features with aesthetic features, the process can be described as follows:

\begin{equation}
\begin{aligned}
y_{i} &= \sigma(MLP_i(f_{i})) \odot c, i \in\mathbf{A},
\end{aligned}
\label{equ:aes}
\end{equation}
where A denotes the set of the attributes extracted:theme, composition, and color. $f_i$ denotes the attributes feature vectors and $\sigma$ denotes the sigmoid function. The $c$ represent the concatenation of all attribute features and aesthetic feature. After that, we concatenate the aesthetic feature and all $y_{i}$ and use a Multilayer Perceptron (MLP) to get the aesthetic score of generated image.

For the Sentiment Alignment part of FVS, we choose a VGG19 \cite{simonyan2014very} model and a RoBERTa model to extract the sentiment feature of generated images and textual prompt respectively. The final dimension of image sentiment feature vectors is 4096 and we additionally train a mapper to harmonize the text sentiment feature vectors which dimension initially is 768.

For the Aesthetic Alignment part of FVS, we use the CLIP \cite{radford2021learning}with ViT-L/14  as the encoder.

\section{More details of adapting models}
\label{B}
\noindent\textbf{More details of data preparation.}
As in the main body in the paper, the entire process of construct training data is illustrated in the Fig.~\ref{fig:collectdata}. For example, we input a prompt shown in the Fig.~\ref{fig:collectdata}, then the Stable Diffusion returns four generated images about the textual prompt. We use FVS to assess the quality score of every four images and reserve the image with the highest score. After that, we use the img-to-img set to generate four new images based on the reserved image via Stable Diffusion \cite{rombach2022high}. 
The above process will be repeatedly executed until the FVS scores of the newly generated four images are all lower than the highest score of the images filtered out in the previous round.

\noindent\textbf{More detail of input prompts.}
In the Tab.~\ref{tab:promptwords}, the column titled "Feature Class" encompasses the three feature classes we utilized: $Compositional$ $Style$, $Color$ $Style$, and $Sentiment$ $Descriptor$. Correspondingly, the column titled "Detailed Content" provides specific terms associated with each of the three feature classes listed in the first column.

The detail content of the $Composition$ $Style$, $Color$ $Style$ and $Sentiment$ $Descriptor$ are shown in the Tab.~\ref{tab:promptwords}. 
The $Composition$ $Style$ component is derived from thirteen composition types within the CADB \cite{zhang2021image} dataset, while $Color$ $Style$ is extracted from the four most common types found within the aesthetics dataset, such as AADB \cite{kong2016photo} and FlickrStyle \cite{karayev2013recognizing}.
For the sentiment descriptors, they are expanded from the Plutchik’s wheel emotion model \cite{plutchik1984emotions}, which expand the eight common emotion: amusement, anger, awe, contentment, disgust, excitement, fear and sadness. In the table, each row within the detailed content of the sentiment descriptors cell contains three sentiment words, which sequentially correspond to the above eight specified emotion.

\begin{figure}[t]
   \centering
   \includegraphics[width=1\linewidth]{./fig/figcollect.pdf}
  \caption{The basic process and methodology for constructing a training dataset.}
   \label{fig:collectdata}
    \vspace{-8pt}
\end{figure}

\begin{table*}[t]
\caption{The content of aesthetic descriptors and sentiment words used in the prompts.}
\label{tab:promptwords}
 \begin{tabular}{c<{\centering}||p{388px}<{\centering}}
\toprule
\textbf{Feature Class} & \textbf{Detailed Content} \\
\midrule
Composition Style  & \makecell[c]{balance, rule of thirds, center, golden ratio, traingle, horizontal, \\
vertical, diagonal, symmetric, curved, radial, vanishing point, pattern}    \\ 
\midrule
Color Style  & vivid color, color harmony, bright, pastel        \\ 
\midrule
Sentiment Descriptors & 
\makecell[c]{ecstasy, joy, serenity; \\
rage, anger, annoyance; \\
amazement, surprise, distraction; \\
admiration, trust, acceptance; \\
loathing, disgust, boredom; \\
vigilance, anticipation, interest; \\
terror, fear, apprehension; \\
grief, sadness, pensiveness}    \\ 
\bottomrule
\end{tabular}
\end{table*}

\begin{figure*}[t]
   \centering
   \includegraphics[width=1\linewidth]{./fig/s2.pdf}
  \caption{Examples of training data, the images with orange frame are chosen into the training dataset.}
   \label{fig:trainingdata}
\end{figure*}

\begin{figure*}[t]
   \centering
   \includegraphics[width=1\linewidth]{./fig/s3.pdf}
  \caption{Full examples of the images in the main text.}
   \label{fig:addition}
\end{figure*}

\begin{figure*}[ht!]
   \centering
   \includegraphics[width=1\linewidth]{./fig/s4.pdf}
  \caption{Additional examples of images generated by SD v1.4 and the adapted one.}
   \label{fig:visual}
\end{figure*}

\noindent\textbf{More examples of training data.}
As shown in Fig.~\ref{fig:trainingdata}, we show some examples of the training data filtered by FVS. 
For each prompt, we concurrently display the images filtered out with the highest and lowest FVS ratings correspondingly.

It can be observed that, for the images with the highest FVS scores, compared to another image generated under the same prompt, they align more closely with the description of the prompt while maintaining a high aesthetic quality. For example, as shown in the Fig.~\ref{fig:trainingdata}.(a), both images convey a sense of sadness, which is consistent with the description in the prompt. However, unlike the first image, the airplane in the selected image aligns more closely with a realistic scenario.
Another example is Fig.~\ref{fig:trainingdata}.(d), compared with the image with lower FVS score, the city in the chosen image exhibits a greater level of detail, and its lighting and overall style more effectively capture the content encompassed in the prompt.

\noindent\textbf{Implementation of adapting model.}
Through FVS, we construct a training set contains 6620 images with higher FVS score.
We fine-tune the UNet of the Stable Diffusion via LoRA \cite{hu2021lora}.
We employ a Adam optimizer with a batch size of 2 due to the constraints of hardware conditions. 
The learning rate, weight decay, and lora rank are set to $1e^{-5}$, $2e^{-2}$, and 4 respectively.
The weight is trained for 15k iterations and the whole process spends about 3 hours.
In the process of generating images, we set the inference step as 50, and for every image with the same prompt, we use the same manual random seed.

\section{More details of adapted model}
\label{C}

\noindent\textbf{More analysis of the correlation between FVS and CLIP Score.}
we further analyze the examples provided in the section on the correlation between FVS and CLIP Score \cite{hessel2021clipscore} in the main text. 
As the image with orange frame, the image, while fulfilling the primary content of a female portrait, more intricately displays the details of the portrait, and overall aligns with the descriptors in the prompt such as `head in focus' and `highly detailed'.
In contrast, as shown in the image with green frame, while the subject of the image satisfies the elements of the prompt, such as `35 mm photo, 99 Rolls-Royce Boat Tail, and model', the overall background of the image fails to reflect `Montreal city', and the lighting cannot be described as `epic'. Thus, the image get a high CLIP score, but received a low FVS score.

\noindent\textbf{More visual results of the adapted model.}
We present a complete version of the examples in the main text, as illustrated in the Fig.~\ref{fig:addition}. 
As observed, the images got from the adapted Stable Diffusion model not only possess enhanced visual appeal compared to those before adapting, but also exhibit greater congruence with the content of the prompt in terms of subject matter, aesthetic descriptors, and overall sentiment information.
For instance, as in the Fig.~\ref{fig:addition}.(d), the image got from the adapted model adhere to the descriptive content specified in the prompt and  create an atmosphere more effectively that aligns with the overall sentiment tone of the prompt simultaneously. 

To further substantiate the role of FVS in filtering high-quality images and its correlation with image quality, we present additional examples of generated images as illustrated in the Fig.~\ref{fig:visual}. The visual results further proved the aforementioned conclusion.

\begin{table}[t]
\small
\caption{The specific content of aesthetic descriptors and sentiment words used in the prompts.}
\label{tab:quant}
 \begin{tabular}{l{16px}<{\centering}c{16px}<{\centering}c{16px}<{\centering}}
\toprule
Generated Model   &  Average FVS $\uparrow$ &Average CLIP Score $\uparrow$\\ 
\midrule
Stable Diffusion v1.4  & 4.3790    &   26.5595  \\ 

Adapted Model(ours) & 4.3895  &26.7342 \\ 
\bottomrule
\end{tabular}
\end{table}

\noindent\textbf{Quantitative analysis of the adapted model.}
We employed the same batch of prompts, which were not utilized in the image filtering process, to generate images with both the Stable Diffusion v1.4 and adapted model, using identical random seeds. Subsequently, we tested the generated images for their average FVS score and CLIP Score \cite{hessel2021clipscore}. The results presented in the Tab.~\ref{tab:quant}.

The results of the quantitative experiment indicate that images filtered using the FVS indeed effectively enhance the quality of the generative model and the high correlation between FVS and image quality.

\small{
\bibliographystyle{IEEEbib}
\bibliography{total}
}